\documentclass[conference]{IEEEtran}
\IEEEoverridecommandlockouts
\usepackage{cite}
\usepackage{amsmath,amssymb,amsfonts}
\usepackage{algorithmic}
\usepackage{graphicx}
\usepackage{textcomp}
\usepackage{xcolor}
\usepackage{svg}
\usepackage{hyperref}
\usepackage{booktabs}

\def\BibTeX{{\rm B\kern-.05em{\sc i\kern-.025em b}\kern-.08em
    T\kern-.1667em\lower.7ex\hbox{E}\kern-.125emX}}

\newcommand{\ie}{{\em i.e.}, }
\newcommand{\eg}{{\em e.g.}, }

\begin{document}

\setlength{\abovecaptionskip}{0ex}
\setlength{\belowcaptionskip}{1ex}
\setlength{\floatsep}{2ex}
\setlength{\textfloatsep}{1.8ex}
\setlength{\intextsep}{0ex}
\setlength{\abovedisplayskip}{0pt}
\setlength{\belowdisplayskip}{0pt}

\title{Modeling Analog-Digital-Converter Energy and Area for Compute-In-Memory Accelerator Design\\
}

\author{\IEEEauthorblockN{Tanner Andrulis}
\IEEEauthorblockA{\textit{MIT} \\ Cambridge, USA \\ andrulis@mit.edu}
\and
\IEEEauthorblockN{Ruicong Chen}
\IEEEauthorblockA{\textit{MIT} \\ Cambridge, USA \\ raychen@alum.mit.edu
}
\and
\IEEEauthorblockN{Hae-Seung Lee}
\IEEEauthorblockA{\textit{MIT} \\ Cambridge, USA \\ hslee@mtl.mit.edu}
\and
\IEEEauthorblockN{Joel S. Emer}
\IEEEauthorblockA{\textit{MIT, Nvidia} \\ Cambridge, USA \\ jsemer@mit.edu}
\and
\IEEEauthorblockN{Vivienne Sze}
\IEEEauthorblockA{\textit{MIT} \\ Cambridge, USA \\ sze@mit.edu}

}

\maketitle

\begin{abstract}
Analog Compute-in-Memory (CiM) accelerators use analog-digital converters (ADCs) to read the analog values that they compute. ADCs can consume significant energy and area, so architecture-level ADC decisions such as ADC resolution or number of ADCs can significantly impact overall CiM accelerator energy and area. Therefore, modeling how architecture-level decisions affect ADC energy and area is critical for performing architecture-level design space exploration of CiM accelerators.

This work presents an open-source architecture-level model to estimate ADC energy and area. To enable fast design space exploration, the model uses only architecture-level attributes while abstracting circuit-level details. Our model enables researchers to quickly and easily model key architecture-level tradeoffs in accelerators that use ADCs.

\end{abstract}


\section{Introduction}

Analog Compute-In-Memory (CiM) accelerators reduce energy in part by using low-energy and high-throughput analog computation. To read computed analog values, CiM accelerators often use analog-digital converters (ADCs). ADCs can consume significant energy and area, and architecture-level decisions such as ADC resolution or number of ADCs can impact the ADC energy and area by orders-of-magnitude~\cite{adc_survey}. For this reason, architecture-level ADC decisions can significantly affect full-accelerator energy and area~\cite{isaac,forms,raella}.

ADC energy and area are key limiting factors in some architectures~\cite{isaac,forms}, and ADC-focused tradeoffs are key optimization targets in other architectures~\cite{raella,cascade,timely,tinyadc,saturation_rram,row_oversubscription}.
Unfortunately, without an architecture-level model of ADC energy and area, prior work is limited to using particular ADC design points (\eg 7-bit, 32nm, $10^9$ converts/second) and can not interpolate (\eg 7-bit, 65nm, vary throughput from $10^6$ to $10^9$ converts per second). The inability to interpolate limits the design points that may be evaluated and precludes exploration of the effects of architectural decisions on ADC energy and area.

To address these challenges, we present an architecture-level model that estimates ADC energy and area given high-level architectural parameters of resolution, throughput, and number of ADCs. In doing so, this model lets users \textbf{(1)} see how architectural decisions influence ADCs, \textbf{(2)} use consistent ADC characterizations to fairly compare architectures, and \textbf{(3)} explore CiM accelerator designs using different ADCs.

We open-source\footnote{\label{opensource}The model and documentation are available at 
\href{https://github.com/accelergy-project/accelergy-adc-plug-in}{https://github.com/accelergy-project/accelergy-adc-plug-in}.} our model, integrating it into the published CiM modeling tool CiMLoop~\cite{cimloop} and using it to model CiM architecture RAELLA~\cite{raella}. 

\section{The Energy and Area Model}

\begin{figure}[]
    \includegraphics[width=\linewidth, keepaspectratio] {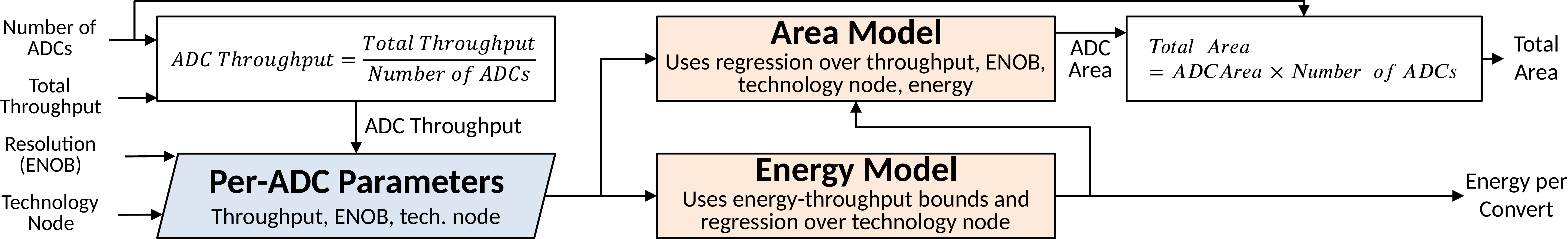}
    \caption{\label{fig:pipeline} The energy and area modeling pipeline.}
\end{figure}

Fig.~\ref{fig:pipeline} shows the modeling pipeline. The model uses four parameters as input: \textbf{(1)} number of ADCs operating in parallel, \textbf{(2)} total throughput (\ie the aggregate number of values that the ADCs convert per second), \textbf{(3)} technology node, and \textbf{(4)} ADC resolution measured as the effective number of bits (ENOB), which measures effective ADC resolution after considering nonidealities such as noise and nonlinearity~\cite{enob}. 

The model uses the total throughput and number of ADCs to calculate per-ADC throughput, then uses per-ADC parameters to calculate per-ADC energy and area. Energy estimates from the energy model are also used as input to the area model.

The energy and area models are generated using statistical analysis of published ADCs. Specifically, energy and area trends~\cite{ad_perf_evolution,ad_perf_evolution_2,murmann_d_2008,jonsson2011empirical,sundstrom_power_2009,race_for_extra_decibel,vogels_architectural_2003,jonsson_area_2011, verhelst_area_2012} are modeled with piecewise power functions that are fit to the Murmann ADC dataset~\cite{adc_survey} using regression$^{\ref{opensource}}$.

Figs.~\ref{fig:energy_tradeoff} and~\ref{fig:area_tradeoff} show the energy and area, respectively, of published ADCs alongside predictions by our model. For ease of visualization, we \textbf{(1)} scale published ADCs to 32nm and set the model to estimate 32nm ADCs, \textbf{(2)} show only lines for 4b, 8b, and 12b ADCs, rounding published ADC ENOB to the nearest of these, and \textbf{(3)} show only ADCs that are near Pareto-optimal. 

We note that the area and energy of published ADCs can vary by orders-of-magnitude even for ADCs with the same architecture-level parameters. This is because each ADC is designed with a different microarchitecture and with different design goals. This model is not intended to capture every variable that may influence ADC area and energy. Rather, the model captures ADC area and energy trends and uses them to estimate a reasonable lower-bound ADC area and energy for a given architecture. 

To model a particular ADC, users may tune the tool's estimated area and energy to match that of the ADC of interest. Users may then use the tool to estimate how the area and energy of that ADC would change given a change in throughput, ENOB, or technology node.

\subsection{The Energy Model}
\label{ssec:energy_model}
\begin{figure}[]
    \includegraphics[width=\linewidth, keepaspectratio] {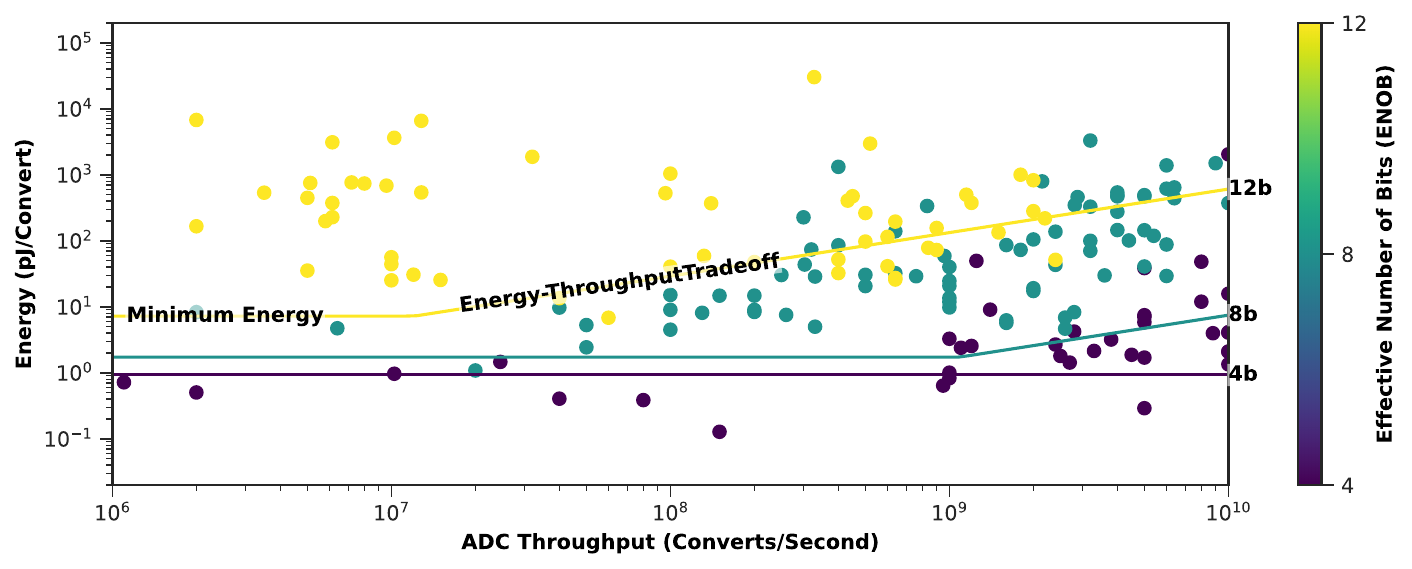}
    \caption{\label{fig:energy_tradeoff} Published ADC throughput versus energy. Lines show energy bounds identified by the model and dots show published ADCs. ADC energy is limited by two bounds that are a function of throughput, ENOB, and technology node.}
\end{figure}

To estimate best-case ADC energy, we use Murmann's observation that ADC energy is limited by two throughput-dependent bounds~\cite{race_for_extra_decibel}. We observe that ADC energy also depends on ENOB and technology node, so we extend Murmann's idea by using best-case energy bounds that are a function of throughput, ENOB, and technology node. At low throughputs, energy is fixed at the \textbf{minimum-energy bound}, shown in the horizontal lines in Fig.~\ref{fig:energy_tradeoff}. At high throughputs, the \textbf{energy-throughput-tradeoff bound} leads to increasing energy with throughput. The energy-throughput-tradeoff bound begins to affect high-ENOB ADCs at relatively lower throughputs due to the design complexity of high-throughput, high-ENOB ADCs. We also find that energy increases exponentially with ENOB.

\subsection{The Area Model}
Prior works~\cite{jonsson_area_2011, verhelst_area_2012} model ADC area using regression based on technology node, throughput, and ENOB. We found that we could improve the correlation coefficient $r$ from $0.66$ to $0.75$ by using energy in place of ENOB\footnote{We hypothesize that energy improves correlation over ENOB alone because low-area layouts also reduce energy through lower wire capacitance.} (note that ENOB still affects energy since energy depends on ENOB).

\begin{equation}~\label{eq:area} \scriptsize
    Area(um^2)=21.1(Tech(nm))^{1.0}(Throughput)^{0.2}(\frac{Energy\ (pJ)}{Convert})^{0.3}
\end{equation}

\begin{figure}[]
    \includegraphics[width=\linewidth, keepaspectratio] {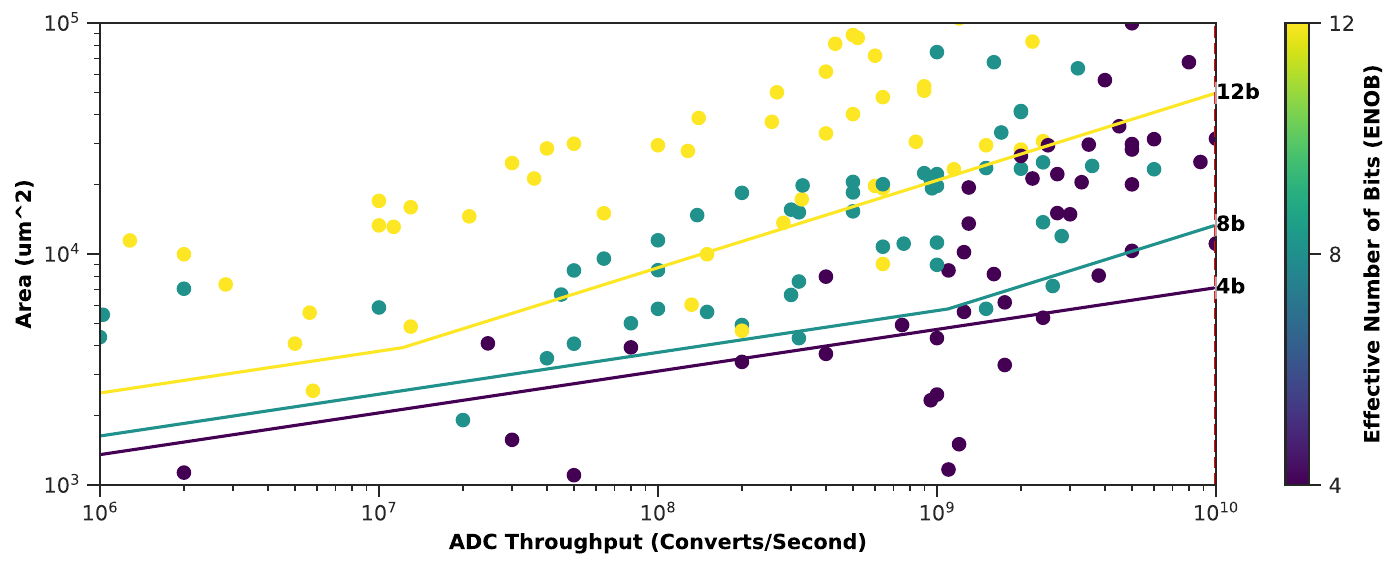}
    \caption{\label{fig:area_tradeoff} Published ADC throughput versus area. Lines show predicted area and dots show published ADCs. As throughput increases, area first increases slowly, then quickly. This is because the two energy bounds influence area.}
\end{figure}

After modeling area with Eq.~\ref{eq:area}, we also optimistically reduce the estimated area to match the lowest-area 10\% of ADCs to predict best-case area. The effects of this model can be seen in Fig.~\ref{fig:area_tradeoff}, which shows published ADC areas alongside predicted area for varying throughputs and ENOBs. The piecewise energy bounds lead to a piecewise relation between throughput and area. We can also see that area increases with throughput and ENOB.

\section{Evaluation}
In this section, use our model to explore how different ADC resolutions, throughputs, and numbers of ADCs affect full-accelerator energy and area. Such explorations are made possible because our model can interpolate between many different design points. Experiments are based on the recent CiM accelerator RAELLA~\cite{raella}.

\subsection{Exploring Architectures while Considering ADCs} If an accelerator performs more computations per ADC convert, it can use fewer ADC converts (less energy), but the additional computations can generate higher-ENOB analog values and require higher-ENOB ADCs (more energy). We explore this trade-off with four parameterizations of RAELLA. Small (S) sums up to $128$ analog values and reads the sum with a 6-bit ADC. Medium (M), large (L), and extra-large (XL) sum up to $512$, $2048$, and $8192$ values, reading results with 7-bit, 8-bit, and 9-bit ADCs, respectively. 

\begin{figure}[]
    \includegraphics[width=\linewidth, keepaspectratio] {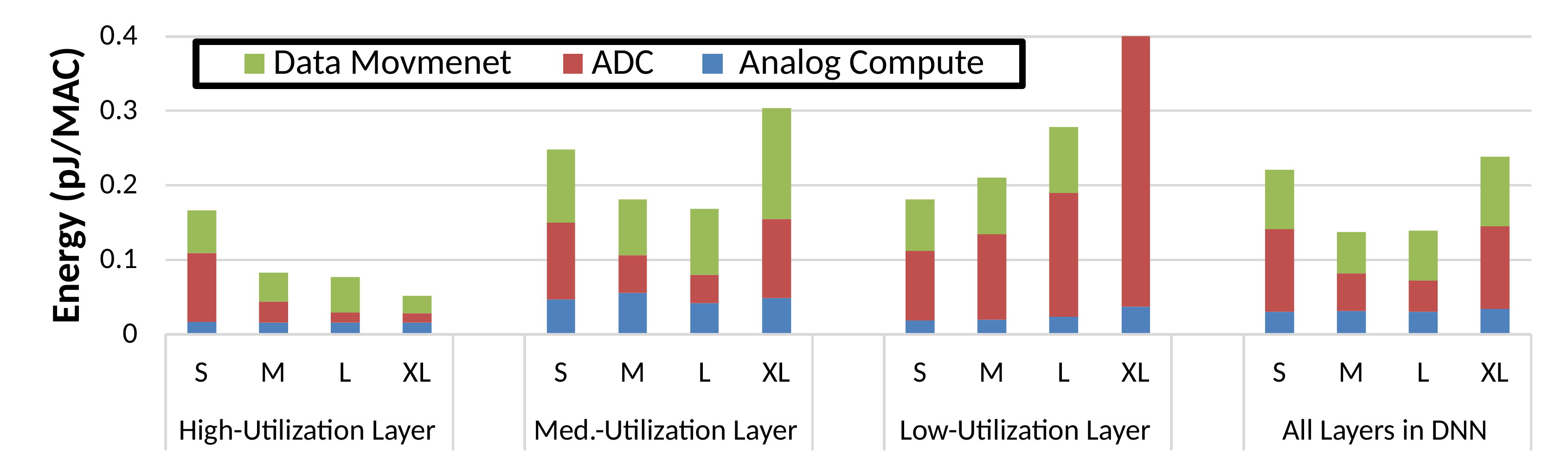}
    \caption{\label{fig:arch_explore} Energy for varying utilization and analog sum size. Summing more analog values and reading the results with higher-ENOB ADCs (towards XL) consumes less energy with higher-utilization DNN layers.}
\end{figure}

Fig.~\ref{fig:arch_explore} shows full-accelerator energy while running layers of varying sizes from the ResNet18~\cite{ResNet} deep neural network (DNN). For the large-tensor layer, summing more analog values reduces ADC energy by performing more computation per ADC convert. For the small-tensor layer, the small tensor size limits the number of values that may be summed, so the architectures with higher-ENOB ADCs consume more energy due to the higher-ENOB ADCs consuming more energy per convert. Over all layers in the DNN, the M and L architectures consume less energy because they balance these two effects.

\subsection{Picking the Number of ADCs for an Architecture}
Using more ADCs consumes more area but also reduces per-ADC throughput, potentially reducing ADC energy. In this test, we generate RAELLA CiM arrays that use 1, 2, 4, 8, and 16 ADCs in parallel. For each configuration, we vary total ADC throughput from ${1.3\times10^9}$ to ${40\times10^9}$ converts per second and measure the overall accelerator energy-area-product (EAP) while running a chosen ResNet18 layer.

\begin{figure}[]
    \includegraphics[width=\linewidth, keepaspectratio] {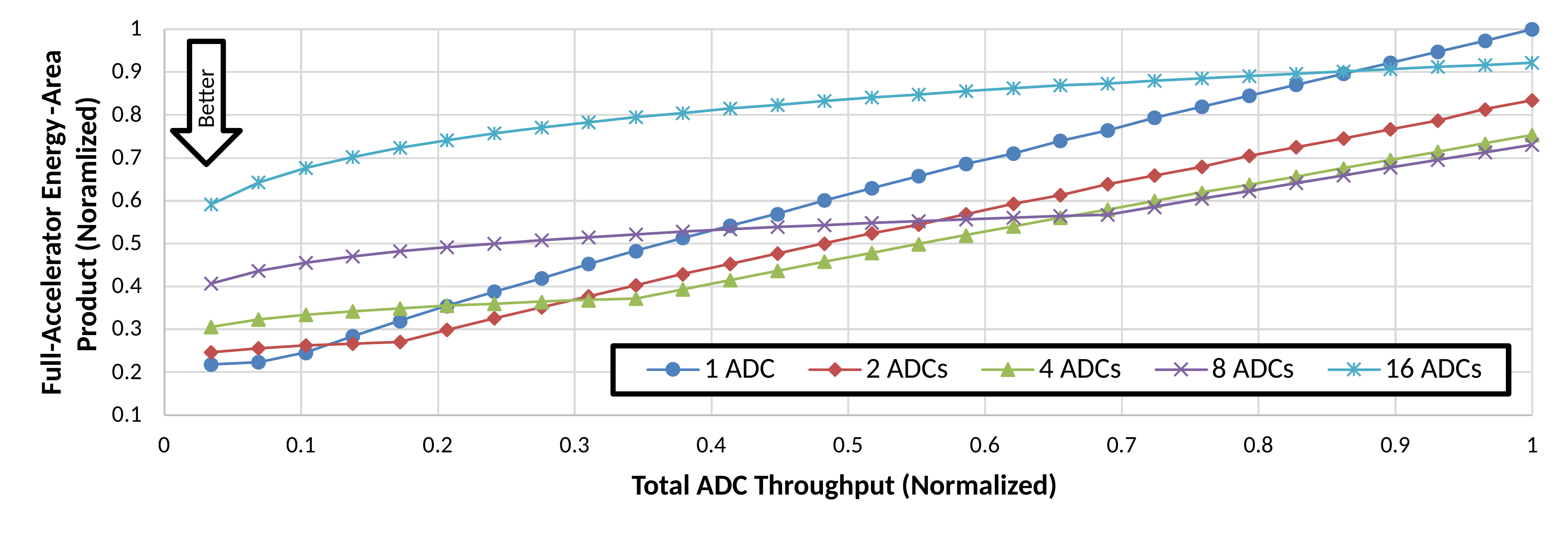}
    \caption{\label{fig:n_adcs_explore} Accelerator energy-area product (EAP) versus number of ADCs for varying throughputs. The number of ADCs significantly impacts EAP, and the lowest-EAP number of ADCs depends on throughput requirements.}
\end{figure}

Fig.~\ref{fig:n_adcs_explore} shows the results. We find that \textbf{(1)} higher total throughput leads to higher EAP as it requires more ADCs or larger, higher-energy ADCs, \textbf{(2)} the choice of number of ADCs can influence overall accelerator EAP by a factor of three, and \textbf{(3)} to minimize EAP, low-throughput accelerators should use fewer ADCs to reduce area and high-throughput accelerators should use more ADCs to reduce energy.

\bibliographystyle{IEEEtran}
\bibliography{main.bib}

\end{document}